# DETERMINATION OF RF SOURCE POWER IN WPSN USING MODULATED BACKSCATTERING

K.Sreedhar[1] and Prof.Y.Sreenivasulu[2]

[1]Department of Electronics and Communication Engineering, VITS (N9)
Karimnagar, Andhra Pradesh, India
sreedhar_kallem@yahoo.com

[2]Head of the Department of Electronics and Communication Engineering, VITS (N9)
Karimnagar, Andhra Pradesh, India
yerraboina@yahoo.com

## *ABSTRACT*

*A wireless sensor network (WSN) is a wireless network consisting of spatially distributed autonomous devices using sensors to cooperatively monitor physical or environmental conditions, such as temperature, sound, vibration, pressure, motion or pollutants, at different locations. During RF transmission energy consumed by critically energy-constrained sensor nodes in a WSN is related to the life time system, but the life time of the system is inversely proportional to the energy consumed by sensor nodes. In that regard, modulated backscattering (MB) is a promising design choice, in which sensor nodes send their data just by switching their antenna impedance and reflecting the incident signal coming from an RF source. Hence wireless passive sensor networks (WPSN) designed to operate using MB do not have the lifetime constraints. In this we are going to investigate the system analytically. To obtain interference-free communication connectivity with the WPSN nodes number of RF sources is determined and analyzed in terms of output power and the transmission frequency of RF sources, network size, RF source and WPSN node characteristics. The results of this paper reveal that communication coverage and RF Source Power can be practically maintained in WPSN through careful selection of design parameters*

## 1. INTRODUCTION

WSN: The development of wireless sensor networks was originally motivated by military applications such as battlefield surveillance [7], [8]. However, wireless sensor networks are now used in many industrial and civilian application areas, including industrial process monitoring and control, machine health monitoring, environment and habitat monitoring, healthcare applications, home automation, and traffic control [1]. In addition to one or more sensors, each node in a sensor network is typically equipped with a radio transceiver or other wireless communications device, a small microcontroller, and an energy source, usually a battery. The envisaged size of a single sensor node can vary from shoebox-sized nodes down to devices the size of grain of dust, although functioning 'motes' of genuine microscopic dimensions have yet to be created. The cost of sensor nodes is similarly variable, ranging from hundreds of pounds few pence, depending on the size of the sensor network and the complexity required of individual sensor nodes. Size and cost constraints on sensor nodes result in corresponding constraints on resources such as energy, memory, computational speed and bandwidth.
The disadvantages of WSN:

- The power unit is a battery.
- The transceiver of a conventional WSN node is typically a short range RF transceiver.
- Compared to the other units of the node, the power consumption of the transceiver is considerably high.

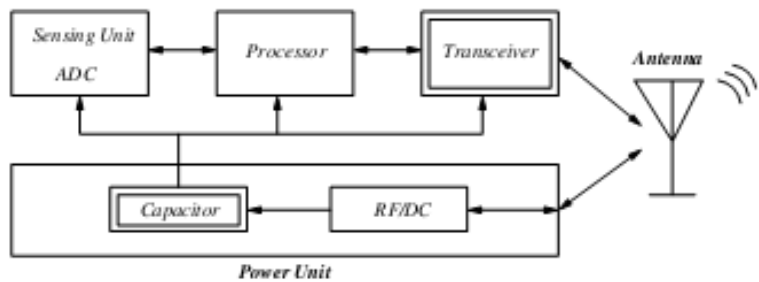

Figure 1.Building blocks of a typical WPSN node.

## 1.1. Module Separation

Our work involves three modules,

*Module 1*: Deriving equation for $P_r$ & $P_t$

*Module 2*: Calculating RF Source Output Power with respect to $P_t$

*Module 3*: Calculating RF Source Output Power with respect to carrier frequency.

*Module 4*: Calculating RF Source Output Power with respect to the area of the event field.

## 2. MODEL OF WPSN

WPSN WITH MB: Wireless passive sensor network proposed in this study is based on MB. The source of energy is an RF power source which is assumed to have unlimited power. The source transmits RF power to run the passive nodes, and it transmits and receives information from WPSN nodes simultaneously. A typical WPSN node hardware is represented in Figure 1. The WPSN node hardware differs from the conventional WSN hardware basically on the power unit and the transceiver. In a conventional WSN node, the power unit is a battery. The power generator, which is an RF to- DC converter is an inherent part of the power unit and is the unique power source of the sensor node. Required power is obtained from the incident RF signal inducing a voltage on the receiver WPSN node. Then, as long as 100mV of voltage is induced on the receiving antenna, RF-to-DC converter yields DC power which is either used to wake up and operate the receiver, sensing and processing circuitries of sensor node, or kept in a charge capacitor to be used later. The transceiver of a conventional WSN node is typically a short range RF transceiver. Compared to the other units of the node, the power consumption of the transceiver is considerably high. For this reason, in WPSN, MB, a passive and less power consuming method is adopted as the main communication mean [3]. Here, the incident signal from the RF source is reflected back by the WPSN node. The node modulates this reflected signal by changing the impedance of its antenna [2], thereby transmits the data gathered from its sensing unit and processed by its processing unit, back to the RF source. The incident signal from the RF source is reflected back by the WPSN node. The node modulates this reflected signal by changing the impedance of its antenna thereby transmits the data gathered from its sensing unit and processed by its processing unit, back to the RF source. The modulated backscattered signal is composed of an information signal, modulated onto a single frequency subcarrier signal, generating a modulated subcarrier signal; this modulated subcarrier signal is then backscatter modulated onto the incoming RF signal.

Advantages of the system are

- Long range communication with the WPSN node is theoretically achievable without increasing the power consumption of the node.

- Combined wireless and battery less operation using the backscatter algorithm

- Cost and effectiveness, systems that use the ultra- high frequencies (UHF) in the industrial, scientific and medical (ISM) radio bands [6].

The transceiver for MB is much less power consuming and fewer complexes, compared to conventional RF transceivers [2]. Furthermore, the maximum communication range of MB is determined by the intensity of the incident signal, and the sensitivity of the corresponding receiver. Thus, long range communication with the WPSN node is theoretically achievable without increasing the power consumption of the node.

In a WPSN deployment, let $P_r$ and $P_t$ be the received power on the passive sensor node and the transmitted power by the RF source, respectively. Then, the RF signal propagates according to Friis' transmission equation [4]

$$P_r = P_t G_t G_r \left( \frac{\lambda}{4\pi R_{rf}} \right)^2 \qquad (1)$$

where $G_t$ and $G_r$ are the antenna gains, $\lambda$ is the wavelength, i.e., the ratio of the speed of light $c$ to the frequency $f$, and $R_{rf}$ is the distance between the RF source and WPSN node. Let the voltage induced on the antenna of WPSN node due to incident signal from RF source be $V_t$. Then, the relation between the received RF power $P_r$ and the induced voltage level $V_t$ is expressed as [4].

$$P_r = \frac{|V_t^2|}{8(R_r + R_l)} \qquad (2)$$

where $R_r$ and $R_l$ are the impedances of the antenna of WPSN node and the RF source, respectively.

According to (1) and (2) and for 4W effective isotropic radiated power (EIRP) output power of RF source, $R_r = R_l = 50$ , $G_t\, G_r = 8.5 dBi$; it is calculated that 100mV can be induced on the antenna of WPSN node from 6.75m at 2GHz, 13.49m at 1GHz, and 26.98m at 500MHz, respectively.
These calculated range values clearly demonstrate that multiple RF sources are needed for the practical implementation of a WPSN deployed over a large event area. Therefore, the required number of RF sources, for a given network size and communication parameters, needs to be determined for sufficient RF coverage, and hence, effective communication in WPSN [12].

## 3. TYPES OF INTERFERENCES AND COMMUNICATION COVERAGE

### 3.1. Source-to-source Interference

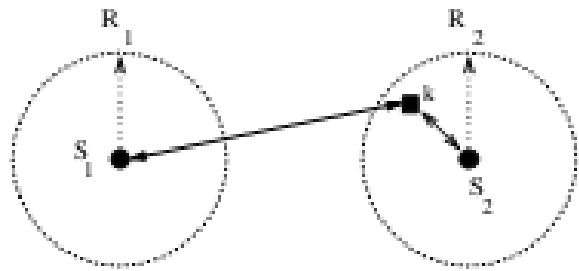

Figure 2. Source–to-source Interference

### 3.2. Source -to-node Interference

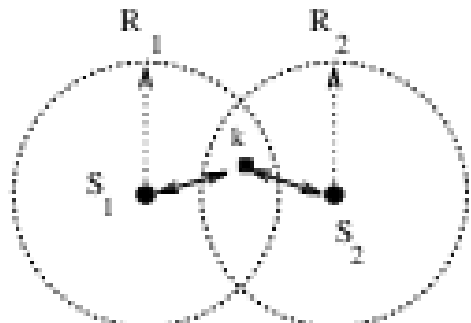

Figure 3. Source- to-node Interference

$N$ sensor nodes are assumed to be randomly distributed over an event area of size    . Communication range of each RF source is represented by a circle of radius $R_{rf}$ . The RF-to-DC converters of WPSN nodes in the range of an RF source are successfully activated by the source, and hence, they are able to reflect the collected data back to the RF source. Here, the ranges of RF sources are considered to be non-overlapping to avoid interference between adjacently deployed RF sources. Source-to-source interference is illustrated in Figure 2. Receiving both the reflected signal from the WPSN node and the strong signal from the source $S_1$ causes interference at source $S_2$. Similarly, source-to-node interference is shown in Figure 3. Communication with two RF sources simultaneously causes interference at the WPSN node $k$. In both cases, communication reliability is hampered due to loss and channel errors. Therefore, in order to avoid these two types of interference, RF sources must have non-overlapping circular ranges of $R_{rf}$ in this analysis. Thus, each passive node is fed by only one RF source in this case. Note that, in fact, if WPSN nodes were fed by multiple RF sources, passive sensor nodes would be able to store more energy in a faster way, and it would be easier for the nodes to receive, store, and transmit power, which would reduce the required number of RF sources for successful communication coverage. Therefore, non-overlapping RF source ranges lead us to the worst-case analysis in this case. Let $k$ be the required number of RF sources to provide MB-based communication coverage over the entire event area of size [2]. Then,

$$k = \frac{\Delta}{\pi R_{rf}^2} \qquad (3)$$

where $R_{rf}$ is the communication range of an RF source.

Substituting (3) for $R_{rf}$ into (1), and then using (2), the required number of RF sources for communication coverage in WPSN, i.e., $k$, can be obtained as

$$k = \frac{2\pi \Delta f^2 |V_t^2|}{c^2 P_t G_t G_r (R_r + R_l)} \qquad (4)$$

where $f$ is the carrier frequency of the RF source, $c$ is the speed of light $c$, i.e., $\lambda = \frac{c}{f}$.

Consequently, (4) can be used to determine appropriate design parameters for effective communication coverage in WPSN [13].

## 4. RESULTS AND NUMERICAL ANALYSIS

Here, the required number of RF sources, i.e., $k$, is investigated for varying event field $\Delta$, RF frequency $f$, output power $P_t$. Note that in order to minimize the overall energy consumption in WPSN, the output power of RF sources needs to be minimized. In this case, for the minimum output power which is just sufficient to induce the necessary voltage, i.e., $V_{t_{min}} = 100mV$, on the receiver of the WPSN nodes, as discussed in Section II, the range of RF sources will be minimum. Therefore, for the worst-case analysis $V_t$ is set to be 100mV. Unless otherwise stated, the remaining simulation parameters are $\Delta = 4x10^{-2} km^2$, $R_r = R_l = 50\Omega$, $G_t G_r = 8.5 dBi$, and $c = 3\times10^8 \, m/s$ [3], [4].

### 4.1. RF Source Output Power

Increasing the RF output power Pt means increasing the range $R_{rf}$ as in (1). An event field can be covered by a smaller number of RF sources if the communication range of RF sources is increased. In Fig. 3(a), $k$ decreases with increasing $P_t$, and hence, increasing $R_{rf}$ range. Moreover, this shows that $k$ increases with carrier frequency for a specific $P_t$ value. This is because WPSN nodes use more energy from RF sources when the communication rate is increased. In Figure 4 shows $P_r$ received on passive sensor node at different induced voltages on received antenna.

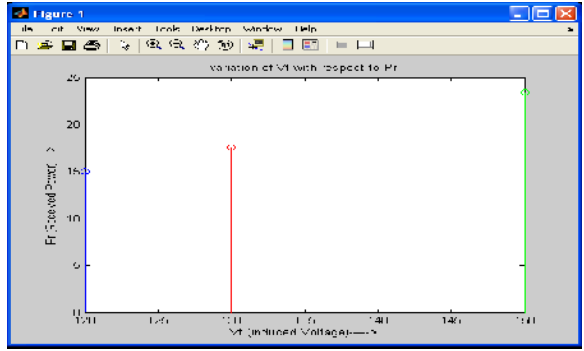

Figure 4. Variation of $V_t$ with respect to $P_r$

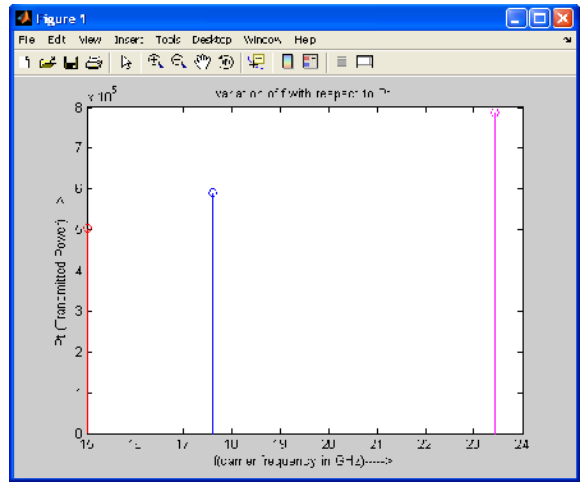

Figure 5. Variation of 'RF' with respect to $P_t$

In Figure 5 shows the $P_t$ transmitted by RF sources at different carrier frequencies.

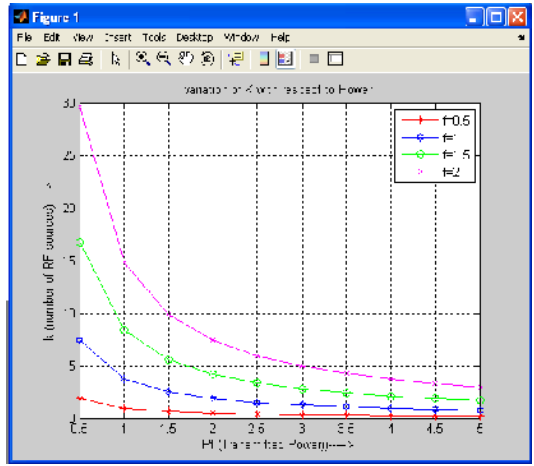

Figure 6. Variation of $k$ with respect to $P_t$

In Figure 6 shows $k$ (No of sources) decreases with increasing $P_t$ (transmitted power by RF sources) and hence increasing $R_{rf}$ range.

## 4.2. Carrier Frequency

As in Figure 5), for a given network dimension and RF out-put power, increasing carrier frequency mandates an increase in the number of RF sources. This is mainly because WPSN nodes become able to use a higher data switching frequency, hence a higher data rate, and the energy consumption for data communication increases. Furthermore, k can be reduced by increasing the output power at a given RF frequency. When output power is increased, the range of RF sources increases, and they start to transmit with higher energy. As a result, each RF source is able to communicate with more WPSN nodes, and a smaller number of RF sources are required for communication connectivity over the event field. On the other hand, the dependence of the required RF output power on frequency and number of RF sources is illustrated in Figure7. The results show the practical applicability of various carrier frequencies with k, in terms of practical values for RF source output power. With an increased number of RF sources, lower RF output power suffices to cover the event field, hence $k$ decreases. On the other hand, for higher frequencies, RF sources transmit with higher RF output power; because signal fading increases at high frequencies [4], [5] and also data communication is performed at a higher rate. These results and observations show that, for a typical WPSN implemented in an area of $40000 m^2$ and communicating with $f = 1GHz$, only 5 RF sources are required at an output power of 1W, as shown in Figure 7.

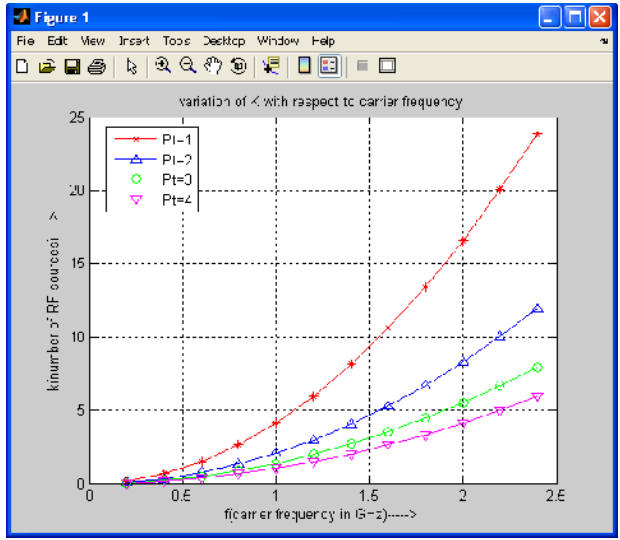

Figure 7.Variation of K with respect to RF

The maximum number of RF sources to achieve communication connectivity in the WPSN increases with RF communication frequency since more RF energy is required for the communication of collected data at a higher rate. Increasing the size of the event field also increases the required number of RF sources for a given output power, because RF sources with a given output power have a limited range determined by their output power, and more such RF sources are needed to cover a larger area. On the other hand, the required number of RF sources decreases with output power. If RF sources transmit at a higher output power, they are able to satisfy (2) for WPSN nodes at a larger distance, hence, their range is increased. This means that the event field can be covered with a smaller number of RF sources [14].

In Figure.7 shows variation of 'k' with respect to carrier frequency. $k$ increases with carrier frequency for a $P_t$ value.

### 4.3. Event Area

As shown in Figure. 5, increasing the network size necessitate communication connectivity over a larger area, and this requires more RF sources, since the range of each RF source is limited by its output power.

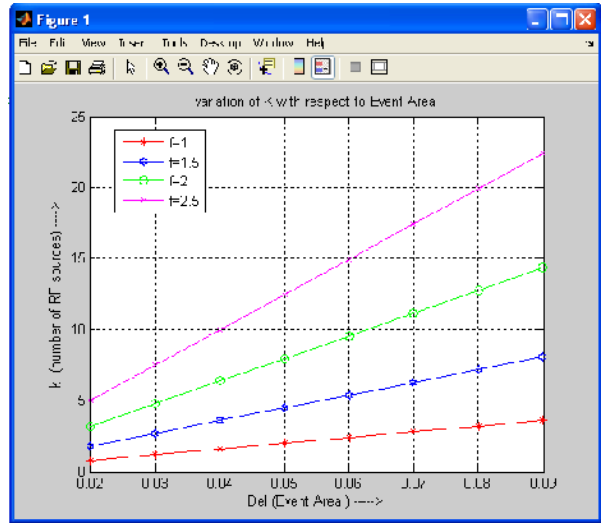

Figure 8. Variation of K with respect to Event Area

In Figure 8 shows variation $k$ with respect to event area increasing the network size necessitates communication connectivity over a large area and this requires no of RF sources since range of each RF source is limited by its output. Consequently, while the results are intuitively not surprising, the analysis shows that the theoretical result in (4) is applicable for the design of practical WPSN deployment cases.

## 5. CONCLUSION

In this paper, RF communication coverage in WPSN is analyzed. The required number of RF sources for effective modulated backscattering-based communication in WPSN is determined in terms of the dimension of the event field, RF communication frequency, and RF output power. The analysis developed here can be used towards determination of design strategies of battery-free WPSN as well as radio frequency identification (RFID) networks. The main focus of this paper is to investigate the communication coverage problem in WPSN. More specifically, minimum number of RF sources to achieve successful MB based communication in WPSN is investigated. Furthermore, the relation between the numbers of RF sources that are required to obtain interference-free RF communication coverage is analyzed in terms of output power and the transmission frequency of RF sources, network size, RF source and WPSN node characteristics.

**Authors**

**K.Sreedhar** received the B.Tech. degree in Electronics and Communication Engineering from JNTUH University, Hyderabad, India in 2005 and M.Tech degree in Communication Systems from JNTUH University, Hyderabad, India in 2009. He attended the International Conference on Technology and Innovation at Chennai. He also attended the National Conference at Coimbatore, Tamilnadu, India on INNOVATIVE IN WIRELESS TECHNOLOGY. He is currently working as an Assistant professor in Electronics and Communication Engineering department in VITS (N9) Karimnagar, Andhra Pradesh, India. He has a Life Member ship in ISTE. He published five International Research papers.

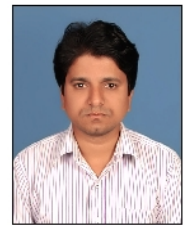

**Professor.Y.Sreenivasulu** received the B.Tech. degree in Electronics and Communication Engineering from Nagarjuna University, Guntur, India and M.Tech degree in ECE from JNTUH University, Hyderabad, India. He worked as an Assistant Professor in Philippians University, Marburg, Germany. He has more than 20 years experience in Teaching Profession. He published a book on Microprocessor and Interfacing. He is currently working as a Head of the department (HOD) in Electronics and Communication Engineering in VITS (N9) Karimnagar, Andhra Pradesh, India. He has a Life Member ship in ISTE. He was pursuing Ph.D in Comparative study of various communication systems in mass media from M.J.P.Rohilkhand University, Bareilly, India.

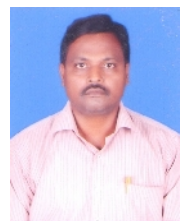